# Thermal Conductivity of Single Wall Carbon Nanotubes: Diameter and Annealing Dependence

M.C. Llaguno, J. Hone[*], A.T. Johnson, J.E. Fischer

*Department of Physics and Astronomy, Laboratory for the Research on the Structure of Matter, University of Pennsylvania, Philadelphia, PA 19104,USA*
[*]*Department of Physics, California Institute of Technology Pasadena, California 91125, USA*

**Abstract.** The thermal conductivity, $\kappa(T)$, of bulk single-wall carbon nanotubes (SWNT's) displays a linear temperature dependence at low T that has been attributed to 1D quantization of phonons. To explore this issue further, we have measured the $\kappa(T)$ of samples with varying average tube diameters. We observe linear $\kappa(T)$ up to higher temperatures in samples with smaller diameters, in agreement with a quantization picture. In addition, we have examined the effect of annealing on $\kappa(T)$. We observe an enhancement in $\kappa(T)$ for annealed samples which we attribute to healing of defects and removal of impurities. These measurements demonstrate how the thermal properties of an SWNT material can be controlled by manipulating its intrinsic nanoscale properties.

## INTRODUCTION

The high thermal conductivity of single wall carbon nanotubes (SWNT's) makes them ideal for thermal management applications. To illustrate, in magnetic field aligned films of SWNT's, a value of 250 W/m-K has been measured at room temperature[1]. For single tubes, theoretical calculations of the thermal conductivity give even higher values of about 10,000 W/m-K[2]. In addition, the unique structure of carbon nanotubes allows for the study of low-dimensional phonons. Specifically, the cylindrical geometry of a tube enforces periodic boundary conditions on the circumferential wave vector, resulting in the formation of 1D phonon 'subbands'- analogous to the electronic subbands. In an isolated SWNT, there are four acoustic phonon modes, with the first optical subband contributing at an energy of a few meV[3]. In heat capacity measurements, the quantized phonon spectrum was observed as a deviation from linear behavior at around 8 K[4] which corresponds to a first subband energy of 4 meV. Similarly, the thermal conductivity of nanotubes should exhibit a linear T-dependence at low T and then a nonlinear trend above a crossover temperature that is directly related to the subband splitting. Previous measurements [5] indeed show a linear $\kappa(T)$ at low T, with an upturn to a higher slope, consistent with this picture. However, the temperature of this upturn (~ 40 K) is much higher than would be expected from the band structure derived from heat capacity measurements. In addition, the behavior of the phonon scattering time $\tau$ is unknown. Therefore it is not possible to conclude with certainty from a single measurement that the observed low-T linear $\kappa(T)$ is due to 1D phonon quantization.

To help determine whether the linear $\kappa(T)$ of bulk SWNTs at low T is due to quantization effects, we have investigated the diameter dependence of $\kappa(T)$ on several bulk samples with different average tube diameters. Because the energy splitting between 1D subbands varies inversely with the radius, we expect that the crossover from linear to non-linear $\kappa(T)$ should occur at higher temperatures in tubes with smaller radii. In agreement with this prediction, the data show increasing crossover temperature with decreasing diameter. In addition, an effect of annealing on $\kappa(T)$ has been observed. Annealed samples exhibit a faster increase in $\kappa(T)$ above the crossover temperature.

## EXPERIMENTAL

The samples were synthesized by laser-ablation at temperatures of 1100 and 1200 C, and purified using acid treatment [6]. Growth temperature is crucial in determining the average tube diameter [6] in that higher temperatures give rise to larger tubes. We measured $\kappa(T)$ for two sets of samples with average diameters of 1.2 and 1.4 nm and comparable degrees of cystallinity and amounts of Ni/Co catalyst impurities. The diameter distribution and degree of crystallinity were determined by TEM and X-ray diffraction (XRD) measurements. Finally, the effects of annealing were examined by vacuum-heating.

We measured the thermal conductivity from 10 to 100 K via a steady state method. Heat was passed through a 50 μm diameter constantan rod of known thermal conductivity and the sample connected in series. The temperature drops across both were measured with differential type E thermocouples. Neglecting radiation losses, which are minimal below 100K, the thermal conductivity of the sample is just the thermal conductivity of the standard multiplied by the ratio of the temperature drop across the standard to that across the sample.

## RESULTS AND DISCUSSION

Fig. 1 shows the normalized data for $\kappa/T$ versus T. All of the samples display linear temperature-dependence at low T, as shown by the constant $\kappa/T$ ratio. The 1.2 nm sample shows an upturn in the slope of $\kappa(T)$ at about 42 K, while the 1.4 nm samples shows an upturn near 35 K. As expected, the larger-diameter sample displays a lower-T upturn. Furthermore, the magnitude of the temperature shift (~15%) is consistent with the relative change in diameter, as predicted by the quantization picture. For each tube diameter, we observed the same behavior in samples having two different morphologies, mat and paper. Therefore, this is evidence that the observed behavior is intrinsic to the tubes and is not dependent on the macroscopic form of the sample.

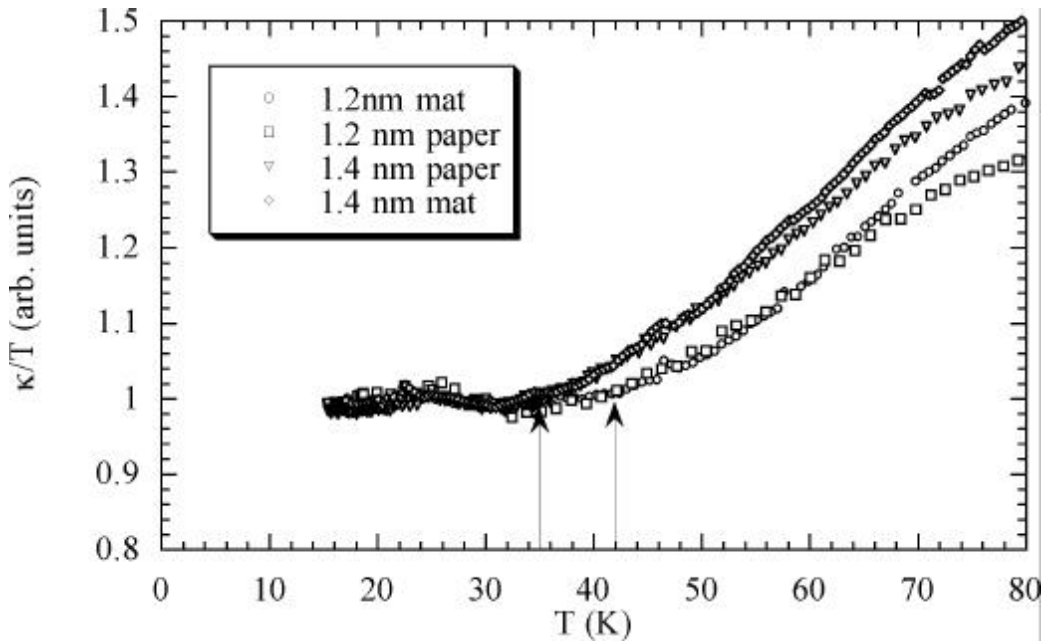

**FIGURE 1.** Normalized data for κ/T. The arrows indicate the two different crossover temperatures at 35K and 42K for the 1.4 nm and 1.2 nm samples, respectively.

In a simple zone-folding model, we calculate the contribution of each phonon subband to the thermal conductivity using the velocities of the four acoustic modes in a nanotube assuming a single scattering time for the phonons [5]. For example, a velocity of 20 km/s and diameter of 1.4 nm give a subband splitting of about 20 meV. Using this value, a calculation of κ(T) for the first phonon subband shows that it should begin to contribute at around 30K. Below this crossover temperature, the linear behavior is due to one dimensional phonons carrying the heat. In Fig. 1, we observe the transition temperature at around the same value, 35 K. However, phonon band structure calculations [3] and heat capacity measurements [4] give subband energies in the range 2.7 meV to 4 meV, much lower than our measured value of around 20 meV, extracted from the model. Since the data provide evidence that the upturn in κ(T) is indeed associated with the population of higher subbands, one possible explanation for the higher cutoff temperature is that the first subband is highly scattering, and we only observe the nonlinear behavior when we access higher phonon subbands.

The limitations of the zone-folding scheme in describing the phonons in nanotubes[3] and the unknown energy dependence of the scattering times necessitate a more detailed modeling of the thermal conductivity in nanotube bundles.

In Fig. 2, we show that annealing leads to a more rapid increase in κ(T) above the crossover temperature. We can correlate this with the XRD data for the magnetic field aligned buckypaper[7]. Upon annealing at 1200 C, rope lattice peaks are evident implying increased crystallinity. With this finding, we can suggest that κ is enhanced at high temperatures due to defect healing, removal of impurities such as $C_{60}$ and improved alignment within individual bundles – effects which are associated with heat treatment. The data suggest that, at low temperatures, κ(T) is less sensitive to intertube scattering and defects than at high T. One other important feature is that upon annealing, the crossover temperature does not change. This is yet another indication that the upturn in κ(T) is intrinsic to the nanotubes.

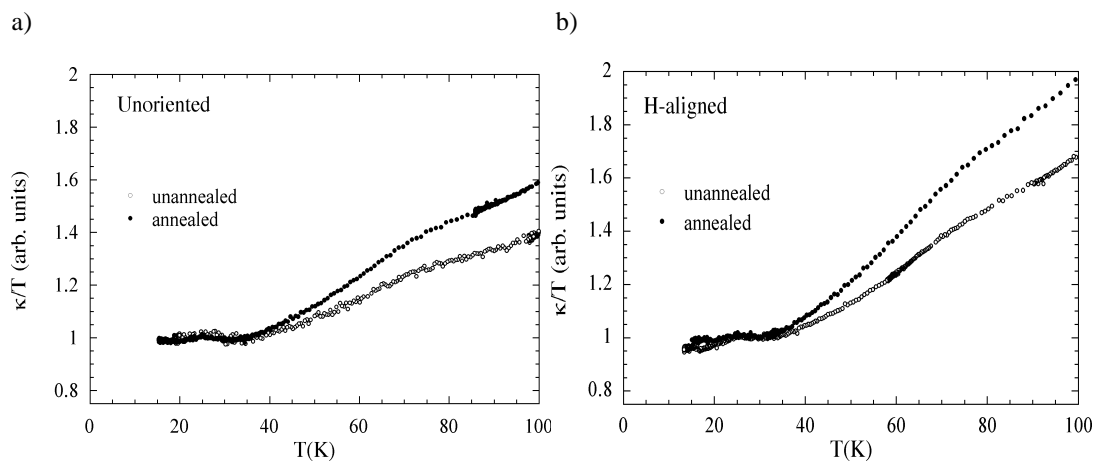

**FIGURE 2.** A faster increase in slope for both a) unoriented and b) magnetic field aligned buckypapers is observed. Note that crossover temperature does not change.

From the measurements, we conclude that we see evidence of phonon quantization in bulk SWNT samples with comparable degrees of crystallinity. This is supported by the diameter dependence of the crossover temperature. Upon annealing at high temperature, we also observe that κ(T) increases faster above the crossover temperature which implies that κ(T) is less sensitive to defects or intertube scattering at low temperatures.

# ACKNOWLEDGEMENTS

This research is supported by the US Department of Energy DOE DEFG02-98ER45701 (MCL and JEF) and by the National Science Foundation ( NSF Grant No. DMR 98-02560) (JH and ATJ).